\documentclass[12,draft]{agujournal2019}
\usepackage{url} 
\usepackage{lineno}
\usepackage{soul}
\usepackage{amsmath}
\usepackage{mathrsfs}

\draftfalse

\usepackage{xcolor}

\newcommand{\la}{\left \langle}
\newcommand{\ra}{\right \rangle}

\newcommand\ve[1]{\boldsymbol{#1}}

\journalname{Journal of Advances in Modeling Earth Systems (JAMES)}

\begin{document}

\title{Benchmarking Turbulence Models to Represent Cloud-Edge Mixing}

\authors{Johannes Kainz\affil{1}, Nikitabahen N. Makwana\affil{2}, Bipin Kumar\affil{2},\\ S. Ravichandran\affil{3}, Johan Fries\affil{4}, Gaetano Sardina\affil{5},\\ Bernhard Mehlig\affil{4,5
}, Fabian Hoffmann\affil{1}}
\affiliation{1}{Ludwig-Maximilians-Universität München, Germany}
\affiliation{2}{Indian Institute of Tropical Meteorology Pune, India}
\affiliation{3}{Centre for Climate Studies, Indian Institute of Technology Bombay, India}
\affiliation{4}{Gothenburg University, Sweden}
\affiliation{5}{Chalmers University of Technology, Sweden}

\correspondingauthor{Fabian Hoffmann}{fa.hoffmann@lmu.de}

\begin{keypoints}
\item Small-scale mixing of a cloudy filament with dry air is simulated using direct numerical simulations and statistical turbulence models.
\item Statistical turbulence models capture the development of thermodynamics during cloud-edge mixing as well as direct numerical simulations.
\item Changes in cloud microphysics are represented correctly only if the statistical models consider the spatial variability of supersaturation.
\end{keypoints}

\begin{abstract}
Considering turbulence is crucial to understanding clouds. However, covering all scales involved in the turbulent mixing of clouds with their environment is computationally challenging, urging the development of simpler models to represent some of the processes involved. By using full direct numerical simulations as a reference, this study compares several statistical approaches for representing small-scale turbulent mixing. All models use a comparable Lagrangian representation of cloud microphysics, and simulate the same cases of cloud-edge mixing, covering different ambient humidities and turbulence intensities. It is demonstrated that all statistical models represent the evolution of thermodynamics successfully, but not all models capture the changes in cloud microphysics (cloud droplet number concentration, droplet mean radius, and spectral width). Implications of these results for using the presented models as subgrid-scale schemes are discussed.
\end{abstract}

\section*{Plain Language Summary}
Although small-scale turbulence is crucial to the development of clouds, the representation of its effects is challenging. Solving the underlying fluid dynamics and cloud microphysical processes on all relevant length- and timescales is possible and highly accurate (so-called direct numerical simulations), but requires enormous computational resources. Therefore, simplified models are used to mimic the effects of small-scale turbulence. Here, we compare four approaches of different complexity to the results from direct numerical simulations. While simpler models capture changes in thermodynamic quantities successfully, the adequate consideration of spatial dependencies is shown to be necessary to represent the development of cloud droplets. 

\section{Introduction}
Today, although considerable progress has been made in understanding the Earth's climate system, the role of clouds still raises questions. One reason for this lack of understanding is that clouds are a multi-scale and multi-process system, ranging from the large-scale organization of cloud fields ($\sim 100\, \mathrm{km}$) to the smallest scales of turbulence ($\sim 1\, \mathrm{mm}$), with which aerosols, cloud droplets, and precipitation interact ($\sim 1-1000\, \mu \mathrm{m}$) \cite{Bodenschatz2010}.

Due to computational constraints, most modeling of the atmosphere focuses on scales larger than 100$\,$m, while smaller-scale processes are parameterized. One of these processes is the mixing of cloudy and cloud-free air \cite{Baker1979}, which can change the cloud microphysical composition and hence the role of clouds in Earth's radiation budget \cite{hoffmann2023small}. The mixing is termed \textit{homogeneous} if the liquid water mixing ratio ($q_\text{c}$) decreases by evaporating all cloud droplets partially, that is, reducing their size but not their number concentration ($N_{\mathrm{c}}$). During extreme \textit{inhomogeneous} mixing, $q_\text{c}$ is reduced by evaporating individual droplets completely, while leaving others unblemished. Thus, $N_{\mathrm{c}}$ decreases, but not the mean droplet size. It is important to note that in both cases, $N_{\mathrm{c}}$ must eventually decrease by the dilution caused by the mixing with entrained air. 

In nature, mixing is usually constrained to the interface of cloudy filaments reaching into subsaturated air. If this filament is not resolved by the model, mixing becomes instantaneous and affects all droplets located in a model grid box simultaneously, which forces mixing to be preferentially homogeneous \cite<e.g.,>{Kainz2023}. Direct numerical simulations (DNSs), which represent all relevant scales of small-scale turbulence, may be applied to study cloud-edge mixing but are constrained to small domains of up to a meter due to the commensurately large computational requirements \cite{Kumar2012,Kumar2013,Kumar2014}. To address larger scales, large-eddy simulations (LESs) are commonly applied, which avoid high spatial and temporal resolution by representing dynamics only on scales larger than tens to hundreds of meters, while parameterizations are used to consider the effects of unresolved scales on the resolved flow \cite{Smagorinsky1963,Deardorff1980}. A key challenge is that unresolved turbulence not only affects the dynamics. Other parameterizations are necessary to represent, for instance,  the effect of turbulent supersaturation fluctuations on the condensational growth of cloud droplets \cite<e.g.,>{HoffmannFeingold2019}.

In this study, we compare different statistical turbulence models and investigate how well they describe the condensation and evaporation of cloud droplets during the mixing of cloudy and cloud-free air. We consider the one-dimensional linear eddy model (LEM) \cite{Kerstein1988}, and three supersaturation-fluctuation models ~\cite{Pope1994}. The first one is the eddy-hopping model (EHM) \cite{Pope1994,Grabowski2017,Abade2018}, which models supersaturation fluctuations as an Ornstein-Uhlenbeck process \cite{ornstein1930theory}. The second model approximates supersaturation fluctuations at the cloud-edge by relaxation to a space-dependent mean (relaxation-to-mean model; RMM) \cite{Pope2000,Fries2021}. The third model rests on the mapping-closure approximation (mapping-closure model; MCM) \cite{Chen1989,Pope1991,Fries2023}. The LEM and EHM have already been used to represent subgrid-scale supersaturation fluctuations in LESs \cite{HoffmannFeingold2019,Chandrakar2021}. \citeA{Fries2021} employed the RMM to estimate the degree of inhomogeneous mixing in the data of \citeA{Beals2015}. The MCM was bench-marked against a DNS of \citeA{Kumar2012}. 
  
Until now, there have been no systematic studies comparing the strengths and weaknesses of the LEM, EHM, RMM, and MCM.
Thus, in the present study we 
\begin{itemize}
    \item  compare all models mentioned above against each other, and with a set of new reference DNSs based on \citeA{Kumar2014},
    \item test to which extent the models describe the thermodynamical and cloud microphysical responses to different turbulence intensities and ambient humidities, and
    \item discuss aspects necessary to consider when the different approaches are used as subgrid-scale models.
\end{itemize}

The paper is organized as follows. First, each model is introduced, and the case setups are presented. Then, two sensitivity studies, one assessing the effect of turbulence intensity and one ambient humidity, are discussed. Differences between the models are highlighted and the applicability of the models as subgrid-scale schemes is discussed. We then conclude with remarks on future work.

\section{Models}

The main way cloud microphysics and dynamics interact is through supersaturation, which drives the condensation and evaporation of cloud droplets. The relative supersaturation with respect to liquid water can be defined as 
\begin{align}
\label{supersaturation}
S({\bf x},t)=\frac{q_{\mathrm{v}}({\bf x},t)}{q_{\mathrm{vs}}[T({\bf x},t),p]}-1, 
\end{align}
and is determined by the water vapor mixing ratio $q_\text{v}$ and absolute temperature $T$, where $q_{\mathrm{vs}}$ is the saturation water vapor mixing ratio, and $p$ is the hydrostatic pressure. Fundamentally, $S$ is a field, defined at any point in space $\bf x$ and time $t$. Supersaturation fluctuations are shaped by the advection of $q_\text{v}$ and $T$, as well as sinks and sources, e.g., the condensation and evaporation of cloud droplets.

All models used in this study aim to determine the spatial and temporal fluctuations of $S$. The most fundamental approach is DNS, described first, followed by short introductions to the LEM, EHM, RMM, and MCM, which idealize the underlying processes to different degrees. Note that the following descriptions cannot provide the depth necessary to understand each approach completely. Thus, references are given to guide the interested reader to further information.

\subsection{Direct Numerical Simulation (DNS)}

DNSs are the most accurate approach to investigate the dynamics of fluids \cite{moin1998direct}, and have been adapted to investigate clouds in the last decades \cite<e.g.,>{vaillancourt2001microscopic}. The applied DNS model is described in \citeA{Kumar2014,Kumar2017}. To predict the velocity of air $\bf u$, where bold font indicates a three-dimensional vector, the DNS solves the incompressible, Oberbeck-Boussinesq-approximated Navier-Stokes equations. These are 

\begin{eqnarray}
\frac{\partial{\bf u}}{\partial {t}}  + ({\bf u}\cdot{\bf\nabla}) {\bf u}  = -\frac{1}{\rho_0}{\bf\nabla}p+
\nu{\bf\nabla}^2{\bf u}+{\bf B}+ {\bf f}_{\mathrm{LS}},
\label{euler2}
\end{eqnarray}
where $\bf\nabla$ and $\partial/\partial t$ denote spatial and temporal derivatives, $\rho_0$ a reference air density, and $\nu$ the kinematic viscosity of dry air. $\bf B$ is the buoyancy force \cite<Eq.\,(6) in>{Kumar2014}, and $\bf f_{\mathrm{LS}}$ a large-scale forcing driving the flow in a statistically stationary fashion \cite<Eq.\,(7) in>{Kumar2014}.

The thermodynamic fields $T$ and $q_\text{v}$ obey
\begin{subequations}
\label{dns:Tq}
\begin{eqnarray}
\label{dns:T}
\frac{\partial{T}}{\partial {t}} + {\bf u}\cdot{\bf\nabla} T & = & D_\kappa
{\bf\nabla}^2 T +\frac{l_{\mathrm{v}}}{c_{\mathrm{p}}}C_{\mathrm{d}},
\\
\frac{\partial{q_{\mathrm{v}}}}{\partial {t}} + {\bf u}\cdot{\bf\nabla} q_{\mathrm{v}} & = & D_\text{v}
{\bf\nabla}^2 q_{\mathrm{v}} -C_{\mathrm{d}},
\end{eqnarray}
\end{subequations}
where $D_\kappa $ is the molecular diffusivity of heat, $D_\text{v}$ molecular diffusivity of water vapor, $l_{\mathrm{v}}$ the enthalpy of vaporization, $c_{\mathrm{p}}$ the specific heat of air at constant pressure. Note that Eq.\,(\ref{dns:T}) does not consider the cooling or warming due to vertical motions, as this term is negligible for the limited vertical domain considered in this study. To ease comparison, this term is not considered in the models described below. The effect of condensation and evaporation on $T$ and $q_\text{v}$ in Eqs.\,(\ref{dns:Tq}) is considered by  

\begin{equation}
\label{condensation rate}
   C_{\mathrm{d}}(\mathbf{x},t)=\frac{4}{3} \pi \frac{\rho_{\mathrm{l}}}{\rho_0}\sum_{\alpha}^N \frac{G({\bf x} - {\bf x_{\alpha}};\Delta V)}{\Delta V } \frac{\text{d} r_{\alpha}^3}{\text{d} t}. 
\end{equation}
Here, $x$ is a location vector and $\mathbf{x_{\alpha}} $ the location of a cloud droplet $\alpha$, which radius growth rate by condensation is $\text{d}r_{\alpha}/\text{d}t$. The number of all droplet is given by $N$. $\rho_l$ is the density of liquid water. The spatial kernel $G$ is unity in a volume $\Delta V$  around ${\bf x}$, and zero otherwise. For DNS, $\Delta V$ is identical to the volume of a DNS grid cell.

Cloud droplet dynamics are described in a Lagrangian fashion, as heavy spherical inertial particles \cite{bec2024statistical}. In the Stokes approximation, their equations of motion read
\begin{subequations} 
\begin{eqnarray}\label{eqn:Lag_position}
\frac{\mbox{d}{\bf x}_{\alpha}}{\mbox{d}t}&=&{\bf v}_{\alpha},
\label{lag1}\\
\frac{\mbox{d}{\bf v}_{\alpha}}{\mbox{d}t}&=&\frac{1}{\tau_\text{s}}\left[{\bf u}({\bf x}_{\alpha})-{\bf
v}_{\alpha}\right]-{\bf g},
\label{lag2}
\end{eqnarray}
\end{subequations}
where ${\bf v}_{\alpha}$ is the droplet velocity and ${\bf u}({\bf x}_{\alpha})$ the air velocity at the droplet's location. ${\bf g}$ describes the acceleration due to gravity. The influence of drag is considered by the particle-response time in Stokes approximation, 
$\tau_\text{s}={2\rho_{\mathrm{l}} r_{\alpha}^2}/{(9 \rho_0 \nu)}$, where $r_{\alpha}$ is the radius the considered droplet. 

The change in droplet radius due to evaporation and condensation is modeled as
\begin{equation}
r_{\alpha}\frac{\mbox{d}r_{\alpha}}{\mbox{d}t}= K S({\bf x}_{\alpha},t),
\label{lag3}
\end{equation} 
which is primarily driven by $S$ determined from $T$ and $q_\text{v}$ via Eq.\,(\ref{supersaturation}) at $\bf {x}_{\alpha}$. The growth rate is scaled by $K=1/[\rho_{\mathrm{l}} (K_\text{D}+K_{\kappa})]$, with $K_\text{D}={R_{\mathrm{v}}T}/{[e_{\mathrm{s}}(T) D_\text{v}]}$ considering the transport of water vapor, and  $K_{\kappa}=\left[{l_{\mathrm{v}}}/{(R_{\mathrm{v}}T)}-1\right]{l_{\mathrm{v}}}/{(\kappa T)}$ the conduction of heat. Here, $R_{\mathrm{v}}$ is the gas constant of water vapor, $e_{\mathrm{s}}(T)$ the saturation vapor pressure, and $\kappa$ the thermal conductivity of air.

\subsection{Linear-Eddy Model (LEM)}

The LEM is a statistical turbulence model developed by \citeA{Kerstein1988}, and was introduced to cloud physics by \citeA{krueger1993linear}. The LEM deviates from DNS by utilizing a one-dimensional description of inertial-range turbulence, in terms of an idealized mapping approach. 

As such, the model does not predict the turbulent fluid velocity, but directly prescribes the relocation of scalars as
\begin{subequations}
\label{LEM:adv}
\begin{eqnarray}
      T(x,t+\text{d}t) = T[M(x),t],\\
    q_\text{v}(x,t+\text{d}t) = q_\text{v}[M(x),t].  
\end{eqnarray}
\end{subequations}

The mapping $M$ mimics turbulent stretching and folding, i.e., it steeps gradients in $T$ and $q_\text{v}$, with the exact mathematical treatment shown in, e.g., Eq.\,(10.2) by \citeA{Menon2011}. $M$ is applied with a probability of $\text{d}t/\tau_{\text{LEM}}$, with $\tau_{\text{LEM}}$ the average time between two turbulent eddies mixing the domain. $\tau_{\text{LEM}}$ is determined from inertial range scaling using the outer scale of turbulence $L$ and the kinetic energy dissipation rate $\varepsilon$, as well as the domain size $L_\text{x}$ \cite<Eq.\,(2.4) in>{krueger1993linear}. $M$ places an eddy at a random location in the domain, and its size $l$ is randomly chosen from the spectrum of eddy lengths in the inertial subrange, bounded by $L_{\mathrm{outer}}$ and the Kolmogorov lengthscale $\eta$  \cite<Eq.\,(2.3) in>{krueger1993linear}, which describes the smallest scale of motion in the flow. Below this scale, molecular diffusion is the dominant driver of energy dissipation.

The evolution of $T$ and $q_\text{v}$ by molecular diffusion and condensation is described by
\begin{subequations}
\label{Tqv_lem}
    \begin{eqnarray}        
        &&\hspace*{-5mm}\frac{\partial T}{\partial t} = D_\kappa \frac{\partial^2 T}{\partial^2 x} +\frac{l_{\mathrm{v}}}{c_{\mathrm{p}}} C_{\mathrm{d}} ,\\
        &&\hspace*{-5mm} \frac{\partial q_{\mathrm{v}}}{\partial t} =D_\text{v} \frac{\partial^2 q_{\mathrm{v}}}{\partial^2 x}- C_{\mathrm{d}}.
    \end{eqnarray}
\end{subequations}
Compared with the DNS Eqs.\,(\ref{dns:Tq}), Eqs.\,(\ref{Tqv_lem}) miss the advection term on their left-hand sides, which is considered in the LEM via $M$ in Eqs.\,(\ref{LEM:adv}). Note that although the LEM is one-dimensional by design, its grid boxes are three-dimensional to represent droplet-concentration-dependent processes correctly. Thus, $C_{\mathrm{d}}$ is evaluated as in the DNS with the reference volume matching the volume of a LEM grid box.

It is assumed that cloud droplets move with the surrounding fluid. Thus, their location changes as
\begin{equation}
x_\alpha(t+\text{d}t) = x[M(x_\alpha)],
\end{equation}
utilizing the same mapping considered in Eqs.\,(\ref{LEM:adv}). While this is a simplification the following models heavily rely on, it is not a necessary requirement for the LEM, in which it is  possible to consider sedimentation or inertia effects that could decouple the motion of droplets from the motion of the fluid \cite{krueger2018economical}. These effects are neglected in the present study.

Droplet condensational growth is determined as
\begin{equation}
r_{\alpha}\frac{\mbox{d}r_{\alpha}}{\mbox{d}t}= K S(x_{\alpha}).
\end{equation} 
Similar to DNS, $S$ is determined from $T$ and $q_\text{v}$ via Eq.\,(\ref{supersaturation}) but at $x_{\alpha}$, the droplet's location in one-dimensional space. This constitutes a minor difference to DNS, where $S$ is evaluated in three-dimensional space.

\subsection{Eddy-Hopping Model (EHM)}
The EHM is a statistical model originally suggested by \citeA{Pope1994}, and applied by \citeA{Grabowski2017} and \citeA{Abade2018} to studying the broadening of droplet size distributions. The model was improved by \citeA{Saito2021}, who reduced the number of model variables, while keeping the capability to reproduce the scaling of the model parameters unchanged. In the EHM, the evolution of individual fluid elements is predicted and it is assumed that cloud droplets move with those fluid elements.

First, prognostic equations for the domain-averaged $T$ and $q_\text{v}$ are solved as
\begin{subequations}
    \begin{eqnarray}        
        &&\hspace*{-5mm}\frac{\partial \overline{T} }{\partial t} =  \frac{l_{\mathrm{v}}}{c_{\mathrm{p}}} C_{\mathrm{d}} ,\\
        &&\hspace*{-5mm} \frac{\partial \overline{q_{\mathrm{v}}}}{\partial t} = - C_{\mathrm{d}},
    \end{eqnarray}
\end{subequations}
where $\overline{(..)}$ indicates the domain average. Because $\overline{T}$ and $\overline{q_{\mathrm{v}}}$ are domain-averaged quantities, advection and molecular diffusion are not considered, and the reference volume to determine $C_{\mathrm{d}}$ encloses the entire domain. From $\overline{T}$ and $\overline{q_\text{v}}$, the desired $\overline{S}$ is determined via Eq.\,(\ref{supersaturation}). 

A perturbation from $\overline{S}$ is predicted for Lagrangian fluid elements by 
\begin{equation}\label{ehm_gov}
    \frac{\mathrm{d}S'_{\alpha}}{\mathrm{d}t}=-C_\text{EHM,1} \frac{S'_{\alpha}}{\tau_{\mathrm{L}}}-C_\text{EHM,2} \frac{S'_{\alpha}}{\tau_{\mathrm{p}}}. 
\end{equation}
The first term represents changes in $S'_{\alpha}$ due to turbulent mixing, steered by the integral timescale of turbulence 
\begin{equation}
\nonumber
    \tau_{\mathrm{L}} = \left(\frac{L_{\mathrm{outer}}^2}{\varepsilon}\right)^{{1}/{3}}.
\end{equation}
The second term depicts changes due to condensation or evaporation by $S'_{\alpha}$, and is derived from $C_\text{d}$. The rate of condensation or evaporation is the phase relaxation timescale 
\begin{equation}
\nonumber
   \tau_{\mathrm{p}}= \frac{1}{a_\text{S} a_\text{K} r_{\mathrm{m}} N_\text{c}}, 
\end{equation}
which is determined from the average droplet radius $r_{\mathrm{m}}$, the domain-averaged droplet concentration mixing ratio $N_\text{c}$, with $a_\text{S}=[{1/q_{\text{v}}} + {l^2_\text{v}}/{(c_{\mathrm{p}} R_{\text{v}}T^2)}]$ and $a_\text{K}= 4\pi \rho_l K$. Note that the parameters $C_\text{EHM,1}$ and $C_\text{EHM,2}$ are tuning parameters to fit the model to the DNS results \cite{Saito2021}.

By combining $\overline{S}$ and $S'_{\alpha}$, the growth of droplets is expressed as 
\begin{equation}
    r_{\alpha}\frac{\mathrm{d} r_{\alpha}}{\mathrm{d} t}=K(\overline{S}+S'_{\alpha}).
\end{equation}
Thus, condensational growth is determined by $S'_{\alpha}$, i.e., a quantity immanent to a specific Lagrangian fluid element. This constitutes a fundamental difference to the representation of condensational growth in DNS and LEM, in which $S$ is evaluated at a specific location in space.

\subsection{Relaxation-to-Mean Model (RMM)}

\citeA{Fries2021} derived a one-dimensional statistical model, which thermodynamics are not determined by $T$ and $q_\text{v}$, but $S$ directly. Note that we do not show the non-dimensional equations of \citeA{Fries2021} to ease comparison with the other models presented here.

In the RMM, two types of Lagrangian fluid elements are simulated, those that contain droplets and those that do not. It is assumed that the droplets move with the fluid elements, and their motion is described by the Langevin equation 
\begin{equation}
\label{eq:velMCM}
\mathrm d  u_{\alpha} = -\frac{u_{\alpha}}{\frac{4}{3}\tau_\text{L}/C_0}  \mathrm d t  + \left( \tau_\text{L} \varepsilon C_0 \right)^{1/2} \text{d}W.
\end{equation}
The first term describes the autocorrelation of a fluid element's motion, with $\frac{4}{3}\tau_\text{L}/C_0$ the autocorrelation timescale \cite{Fries2021}, and the parameter $C_0$ according to \citeA{Pope2011}. The second term represents turbulent motion as Brownian increments, with $\text{d}W$ being an increment of a Wiener process with zero mean and variance, d$t$/$\tau_{\mathrm{L}}$, equal to a time step \cite{Pope1991}. $\left( \tau_\text{L} \varepsilon C_0 \right)^{1/2}$ is the standard deviation of the velocity increments. The displacement of a fluid element is given by 
\begin{equation}
\label{eq:posMCM}
\frac{{\rm d} x_{\alpha}}{{\rm d}t} =  u_{\alpha}. 
\end{equation}

Similar to Eq.\,(\ref{ehm_gov}) in the EHM, $S_{\alpha}$ is determined by
\begin{equation}
\label{eq:supsatMCM}
\frac{{\rm d}S_{\alpha}}{{\rm d}t} = - C_\text{RMM,1} \frac{S_{\alpha} -\langle {S}(x_{\alpha},t)\rangle}{\tau_\text{L}} - C_\text{RMM,2} \frac{\langle r(t) S(x_{\alpha},t) \rangle / r_\text{i}}{\tau_\text{p,i}}.
\end{equation}
The first term describes the relaxation of $S_{\alpha}$ to $\langle {S}(x_{\alpha},t)\rangle$, the ensemble-averaged $S$ determined from all simulated fluid elements at $t$ and $x_{\alpha}$ \cite{Fries2021}. The ensemble average is indicated by $\langle ..\rangle$.
The second term describes the effect of condensation and evaporation, and can be derived from $C_\text{d}$. Similar to the first term, $\langle r(t) S(x_{\alpha},t) \rangle$ is an ensemble-average determined from all simulated fluid elements at $t$ and $x_{\alpha}$. Here, $r_\text{i}$ and $\tau_\text{p,i}$ are the initial droplet radius and phase relaxation timescale, respectively. $C_\text{RMM,1}$ and $C_\text{RMM,2}$ are empirical constants. 

Droplet condensational growth is determined as 
\begin{equation}
    r_{\alpha}\frac{{\rm d}r_{\alpha}}{{\rm d}t}  = K S_{\alpha},
\end{equation}
similar to EHM. However, $S_{\alpha}$ is linked to a $x_\alpha$, which enables RMM to consider spatial information of the mixing process that is missing in the EHM. 

\subsection{Mapping-Closure Model (MCM)}
For a better representation of supersaturation fluctuations,  \citeA{Fries2023} developed another statistical mixing model based on the mapping-closure approach by \citeA{Chen1989} and \citeA{Pope1991}.

Similar to the RMM, the MCM is based on replacing $T$ and $q_\text{v}$ with a single equation for $S$,
\begin{equation}
\label{MCM:S}
\frac{\partial{S}}{\partial t} + {\bf u}\cdot{\bf\nabla} S = D
{\bf\nabla}^2 S + \frac{1}{q_\text{vs}(T_0,p_0)} C_{\mathrm{d}},
\end{equation}
where $T_0$ and $p_0$ are reference temperatures and pressures, respectively. To retrieve $S$, MCM does not solve Eq.\,(\ref{MCM:S}) directly, but employs a mapping $X$ from a time-independen Gaussian-distributed random variable $\xi(t)$
, with zero mean and unit variance, to $S(\ve x,t)$, which --- in a statistical sense --- is interpreted as an \textit{arbitrarily} distributed random variable. Accordingly, for a given $\ve x$ and $t$, 
\begin{align}
\label{eq:ansats}
S(\ve x,t) = X\{\xi[\ve x/\lambda_\text{MCM}(t)],t\},
\end{align}
where $\lambda_\text{MCM}(t)$ a time-dependent length scale. By inserting Eq.\,(\ref{eq:ansats}) in (\ref{MCM:S}), with some further considerations detailed in \citeA{Fries2023}, $X$ is predicted as
\begin{equation}\label{mapevol}
\frac{\partial X}{\partial t}= \varphi(t)\biggl(-\delta \frac{\partial X}{\partial \delta }
+\frac{\partial^2 X}{\partial \delta^2 }\biggr)-\frac{\tau_\text{L}}{\tau_\text{p,i}} \la C_{\mathrm{d}}| S = X \ra,
\end{equation}
where $\delta$ is the sample-space variable corresponding to $\xi$, with the sample space containing all possible realizations of $\xi$. Further, $\varphi(t)$ is a relaxation rate, and $\la C_{\mathrm{d}}| S = X \ra$ is a conditional ensemble-averaged condensation rate. $\varphi(t)$ and the aforementioned $\lambda_\text{MCM}(t)$ are \textit{unknown} for most applications. Hence, they are determined from DNS here, while theoretical closures may exist for specific cases. 

Similar to Eqs.\,(\ref{ehm_gov}) and (\ref{eq:supsatMCM}) in EHM and RMM, MCM predicts $\xi_\alpha$ for multiple fluid elements using a Langevin equation, such that
\begin{equation}
\label{eq:ou}
  \mathrm{d} \xi_\alpha = - R(t) \xi_\alpha \, \mathrm{d} t + \sqrt{2 R(t)} \, \mathrm{d} W.
\end{equation}
This approach ensures that the ensemble of $\xi_\alpha$ relaxes to a Gaussian as assumed for $\xi$ above. Moreover, \citeA{Fries2023} set $R(t)= C_\text{MCM} \varphi(t)$, with $C_\text{MCM}$ a fitting parameter from DNS. 

Similar to EHM and RMM, the droplet condensational growth equation is  
\begin{equation}
    r_{\alpha}\frac{{\rm d}r_{\alpha}}{{\rm d}t}  = K X(\xi_\alpha,t).
\end{equation}
Here, however, the mapping $X$ is crucial to determine $S$ from $\xi_\alpha$ via Eq.\,(\ref{eq:ansats}).

\section{Case Setup}

We simulate the mixing of a cloudy filament with dry air. In this section, we will summarize the setup of the simulated cases, and how the different models are initialized.

\subsection{Simulated Cases}

\begin{table}[t]
\caption{\label{tab:supersaturation} Parameters for the initial $T$, $q_\text{v}$, and $S$ profiles.}
\begin{tabular}{ccccccccc}
\hline
 & $T_\text{e}$& $q_{\text{vs,e}}$ &$q_{\text{v,e}}$ & $S_\text{e}$ &  $T_\text{c}$ & $q_{\text{vs,c}}$ & $q_{\text{v,c}}$ & $S_\text{c}$ \\
  &   ($\text{K}$) &   ($\text{g}\,\text{kg}^{-1}$) &  ($\text{g}\,\text{kg}^{-1}$) & ($-$) &   ($\text{K}$) &  ($\text{g}\,\text{kg}^{-1}$) &  ($\text{g}\,\text{kg}^{-1}$) & ($-$) \\\hline
$n_0$& 271.4099& 4.0{48} & 0.031&{-0.9922}&270.7535&3.86{6}&4.015&{0.03852}\\
$n_1$& 271.3074& 4.0{19}& 1.027&{-0.7444}&270.8151&3.88{3}& 4.015&{0.03403}\\
$n_2$&271.2049& 3.99{0}& 2.023&{-0.4929}&270.8768&3.89{9}&4.015&{0.02957}\\
$n_3$&271.1025& 3.96{2}& 3.019&{-0.2379}& 270.9384&3.91{6}&4.015&{0.02513}\\\hline
\end{tabular}
\end{table}

\begin{table}
\caption{\label{tab:tke} Turbulence parameters used in this study.}
\begin{tabular}{llrrrrr}
\hline
$\varepsilon$                &($\text{cm}^2\,\text{s}^{-3}$)\hspace*{5mm}  & 1\hspace*{5mm}  &10\hspace*{5mm} & 33.75\hspace*{5mm} &100 \hspace*{5mm}&1000\hspace*{5mm}\\\hline
 $\text{TKE}$& ($\text{cm}^2\,\text{s}^{-2}$)\hspace*{5mm} &      7\hspace*{5mm} &   34.4\hspace*{5mm}   &  78.8\hspace*{5mm}   & 168.1\hspace*{5mm}    & 799.9\hspace*{5mm}        \\
$\tau_{\mathrm{L}}  $                      &($\text{s}$)\hspace*{5mm} & 7\hspace*{5mm}     &  3.44\hspace*{5mm}    &  2.34\hspace*{5mm}   &    1.68\hspace*{5mm}    &     0.8\hspace*{5mm}    \\\hline
\end{tabular}
\end{table}

All models simulate a domain that is characterized by the lengthscale $L_{\mathrm{x}} = 51.2\, \mathrm{cm}$, irrespective of the dimensionality of the considered approach. All simulations are run for $40\,\text{s}$. Without the aim to simulate a specific cloud type, we prescribe reference values typical for low-level clouds (e.g., shallow cumulus or stratocumulus), with $T_0=271\,$K, $\rho_0=1.06\,\text{kg}\,\text{m}^{-3}$, and $p_0=824.6\,\text{hPa}$.

To represent the mixing of a cloudy filament with dry air, we initialize $q_{\mathrm{v}}$ and $T$ such that they depict the change from dry to cloudy air along each model's primary dimension, following the original setup proposed by \citeA{Kumar2014}. We assume that the model's primary dimension is the $x$-axis. Thus, 
\begin{equation}
\label{qv_init}
    q_\text{v,i}(x)\equiv q_{\mathrm{v}}(x,t=0)=(q_{\mathrm{v,c}}-q_{\mathrm{v,e}})\exp{\left[ -\lambda \left(x-\frac{L_{\mathrm{x}}}{2}\right)^8\right]}+q_{\mathrm{v,e}},
\end{equation}
with $q_{\mathrm{v,e}}$ the minimum of $q_\text{v}$ found in the environment, $q_{\mathrm{v,c}}$ the maximum of $q_{\mathrm{v}}$ inside the cloud, and $\lambda=1.45\times 10^{-10}\,\mathrm{cm}^{-8}$. The $T$ profile is 
\begin{equation}
\label{T_init}
    T_\text{i}(x)\equiv T(x,t=0)=T_0 \left\{1 - \left( \frac{R_{\mathrm{v}}}{R_{\mathrm{d}}} - 1 \right) [q_{\mathrm{v,i}}(x)-q_{\mathrm{v,e}}]\right\},
\end{equation}
where $R_{\mathrm{d}}$ is the gas constant of dry air. The corresponding initial $S_\text{i}$ profile is determined from $q_\text{v,i}$ and $T_\text{i}$ via Eq.\,(\ref{supersaturation}).

To investigate the impact of ambient humidity, we vary $q_\text{v,i}$ by adapting $q_\text{v,e}$ as
\begin{equation}
	q_{\mathrm{v,e}} = q_{\mathrm{v,e}}^{\mathrm{orginal}} + n \frac{q_{\mathrm{v,c}} - q_{\mathrm{v,e}}^{\text{original}} }{4}, \label{eq:initial_profile}
\end{equation}
with $n$ from $0$ to $3$, and $q_{\mathrm{v,e}}^{\text{original}}$ taken from \citeA{Kumar2014}. The corresponding setups are called $n_0$ to $n_3$, respectively. The resultant values for $T$, $q_\text{v}$, $q_\text{vs}$, and $S$ in the environment (subscript e) and in the cloud (subscript c) are summarized in Tab.\,\ref{tab:supersaturation}. To assess the impact of turbulence, $\varepsilon$ is varied between $1$ and $1000\,\text{cm}^2\,\text{s}^{-3}$, with the corresponding turbulence parameters stated in Tab.\,\ref{tab:tke}.

\begin{figure}[h!]
    \centering
    \includegraphics[width=\textwidth]{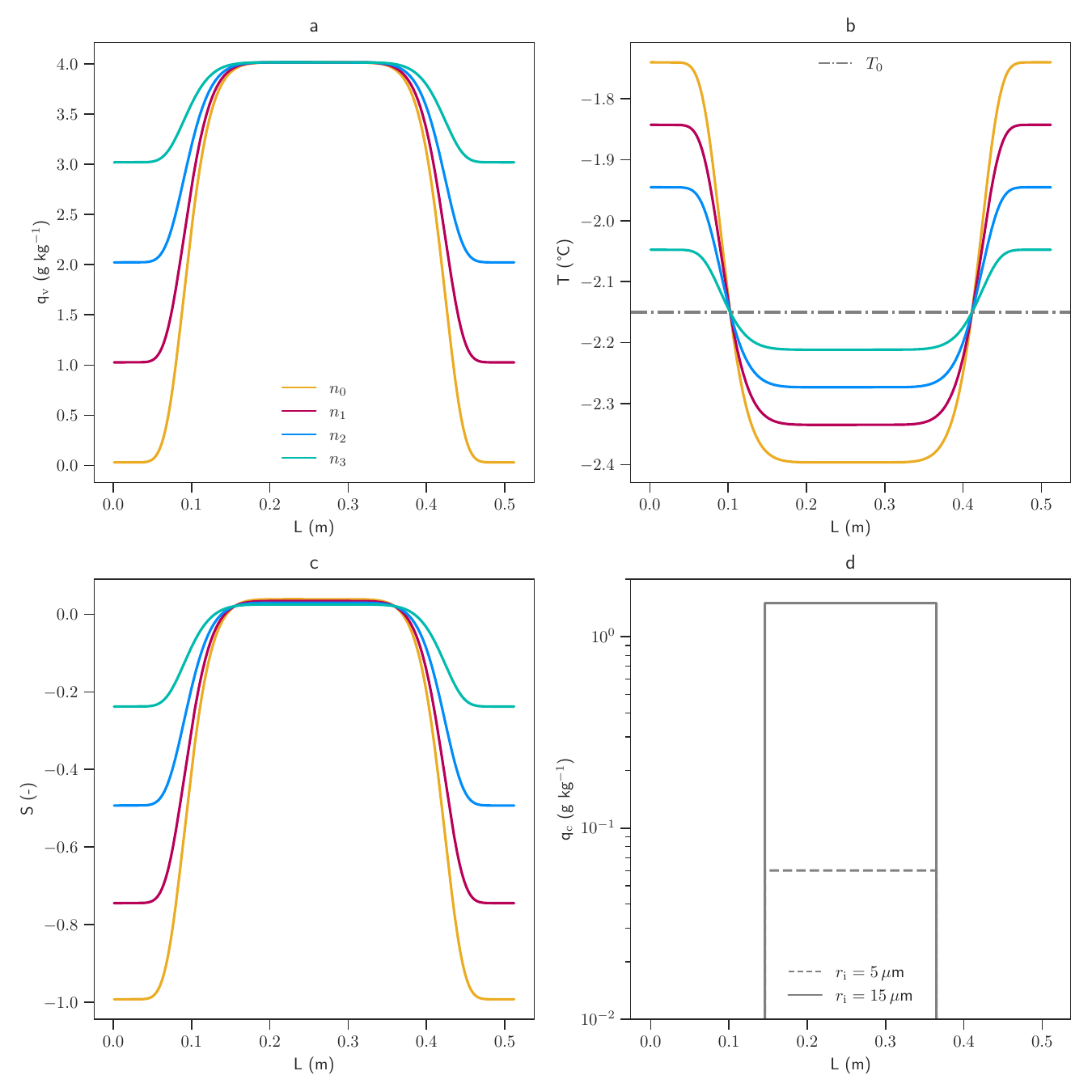}
    \caption{Initial (a) $q_{\mathrm{v,i}}$, (b) $T_\text{i}$, (c) $S_i$, and (d)  $q_{\mathrm{c,i}}$ profiles for different thermodynamics (denoted $n_0$-$n_3$, colored lines) or $r_\text{i}$ (gray lines). In (b), the reference temperature $T_0$ is shown (gray dashed-dotted line).}
    \label{fig:initialprofiles}
\end{figure}

Cloud droplets are initialized as monodisperse particles with initial radii $r_\text{i} = 5$ or $15\,\mu\text{m}$ in the the central $22\,\text{cm}$ of the domain. We refer to this part of the domain as the \textit{cloudy part}. Note that although the cloudy part covers most of the supersaturated domain, these regions are not identical due to the varying ambient humidity. For all cases, the same domain-averaged $N_{\mathrm{c,i}}=47\,\text{mg}^{-1}$ is initialized, corresponding to $109\,\text{mg}^{-1}$ in the cloudy part. This results in $q_\text{c,i}=0.06\,\text{g}\,\text{kg}^{-1}$ or $1.54\,\text{g}\,\text{kg}^{-1}$ for $r_\text{i} = 5$ or $15\,\mu\text{m}$ in the cloud part. Figure~\ref{fig:initialprofiles} summarizes the initial profiles for $q_{\mathrm{v,i}}$, $T_{\mathrm{i}}$, $S_\text{i}$, and $q_\text{c,i}$ for all thermodynamic setups ($n_0$ to $n_3$) and $r_\text{i}$. 

\begin{table}[h!]
\caption{\label{tab:DNS_para} Further parameters used in this study.}
\centering
   \begin{tabular}{p{6cm}ccc}
    \hline
         quantity & symbol & unit & value \\
         \hline
         Kinematic viscosity & $\nu$ & $\text{m}^2\,\text{s}^{-1}$ & $1.5\times10^{-5}$ \\
         Molecular diffusivity of heat & $D_\kappa$ & $\text{m}^2\,\text{s}^{-1}$ & $2.23\times10^{-5}$ \\
         Molecular diffusivity of water vapor & $D_\text{v}$ & $\text{m}^2\,\text{s}^{-1}$ & $2.16\times10^{-5}$ \\
         Enthalpy of vaporization & $l_{\mathrm{v}}$ & $\mathrm{J}\,\mathrm{kg}^{-1}$ & $2.5\times10^{6}$ \\
         Specific heat of air at const. pressure & $c_{\mathrm{p}}$ & $\mathrm{J}\,\mathrm{K}^{-1}\,\mathrm{kg}^{-1}$ & $1005.0$ \\
         Density of liquid water & $\rho_{\mathrm{l}}$ & $\text{kg}\,\text{m}^{-3}$ & $1000.0$ \\
         Gas constant of water vapor & $R_{\mathrm{v}}$ &          $\mathrm{J}\,\mathrm{K}^{-1}\,\mathrm{kg}^{-1}$ & $461.5$ \\
         Gas constant of dry air & $R_{\mathrm{d}}$ & $\mathrm{J}\,\mathrm{K}^{-1}\,\mathrm{kg}^{-1}$& $287.0$ \\
         Thermal conductivity & $\kappa$ & $\mathrm{J}\,\mathrm{m}^{-1}\,\mathrm{s}^{-1}\,\mathrm{K}^{-1}$ & $2.38\times 10^{-2}$ \\
         \hline
    \end{tabular}
\end{table}

All models use the parameters summarized in Tab.\,\ref{tab:DNS_para}. Moreover, to ease comparability, all models use the same
\begin{eqnarray}
          q_{\mathrm{vs}}(T,p) =\frac{R_\text{d}}{R_\text{v}}\frac{e_\text{s}(T)}{p-e_\text{s}(T)} \approx \frac{R_\text{d}}{R_\text{v}} \frac{e_\text{s}(T)}{p}, 
\end{eqnarray}
with 
\begin{equation}
    e_\text{s}(T) = b_{\text{s,1}} \exp{\left(-\frac{b_{\text{s,2}}}{T}\right)},
\end{equation}
and the parameters $b_{\text{s,1}}=2.53\times 10^8\,\mathrm{kPa}$ and $b_{\text{s,2}}=5420\,\mathrm{K}$ from \citeA{Rogers}.

\subsection{Model Initialization}

In DNS and LEM, the initial distributions of $q_\text{v,i}$, $T_\text{i}$, and hence $S_\text{i}$ are represented on a numerical grid with resolutions of $1.0$ and $0.1\,\text{mm}$, and $134\,217\,728$ and $5\,120$ grid boxes, respectively. In the initially cloudy part of the domain, the approaches initialize 6\,771\,519 and 2\,594 Lagrangian cloud droplets at random $\textbf{x}_{\alpha\text{,i}}$ and  $x_{\alpha\text{,i}}$, respectively. For DNS and LEM, $r_{\alpha,\text{i}} = r_\text{i}$ for all simulated droplets initially. 

For EHM, $\overline{S_\text{i}}$ is determined from $\overline{T_\text{i}}$ and $\overline{q_\text{v,i}}$. $S'_{\alpha,\text{i}}$ is calculated such that $\overline{S_\text{i}}+S'_{\alpha,\text{i}}=S_\text{i}(x_\text{i})$, where $x_\text{i}$ are randomly picked locations in the cloudy region. (Note that $x_\text{i}$ is only relevant during the initialization, and is not considered in subsequent calculations of the EHM.) In total, 2,594 Lagrangian cloud droplets are simulated, with their initial $r_{\alpha,\text{i}} = r_\text{i}$. 

RMM assigns $S_{\alpha,\text{i}}=S_\text{i}(x_{\alpha,\text{i}})$ to Lagrangian fluid elements at random locations $x_{\alpha,\text{i}}$. The fluid elements are of two types, representing air with or without droplets, respectively. $1.4\times 10^5$ air elements are initialized in the entire domain, $2\times 10^5$ droplet elements within the cloudy part. The latter are assigned the radius $r_{\alpha,\text{i}} = r_\text{i}$, while the prior $r_{\alpha,\text{i}} = 0$ throughout the simulation.

MCM determines the initial mapping $X_\text{i}$ from the relation $F_\text{i}[X_\text{i}(\delta)] = \Gamma(\delta)$, where $F_\text{i}(S)$ is the initial cumulative distribution function (CDF) of $S$ from DNS, and $\Gamma(...)$ is the CDF of a standardized Gaussian-distributed variable. One proceeds by sampling $10^6$  fluid elements with $\xi_{\alpha,\text{i}}$ from a standardized Gaussian conditional on that they are mapped to a supersaturation $X_\text{i}(\xi_{\alpha,\text{i}})$ found within the droplet-containing region of the initially cloud part of the domain. The fluid elements are assigned the initial radius $r_{\alpha,\text{i}} = r_\text{i}$.

Simulations using the LEM and EHM are repeated each 1000 times to gain reliable ensemble averages. Because RMM and MCM compute a much larger number of fluid elements, ensemble-averaging is not necessary. Also for DNS, results from only one simulation are considered.

\section{Results}

\subsection{General Development}

\begin{figure}[h!]
    \centering
    \includegraphics[width=\textwidth]{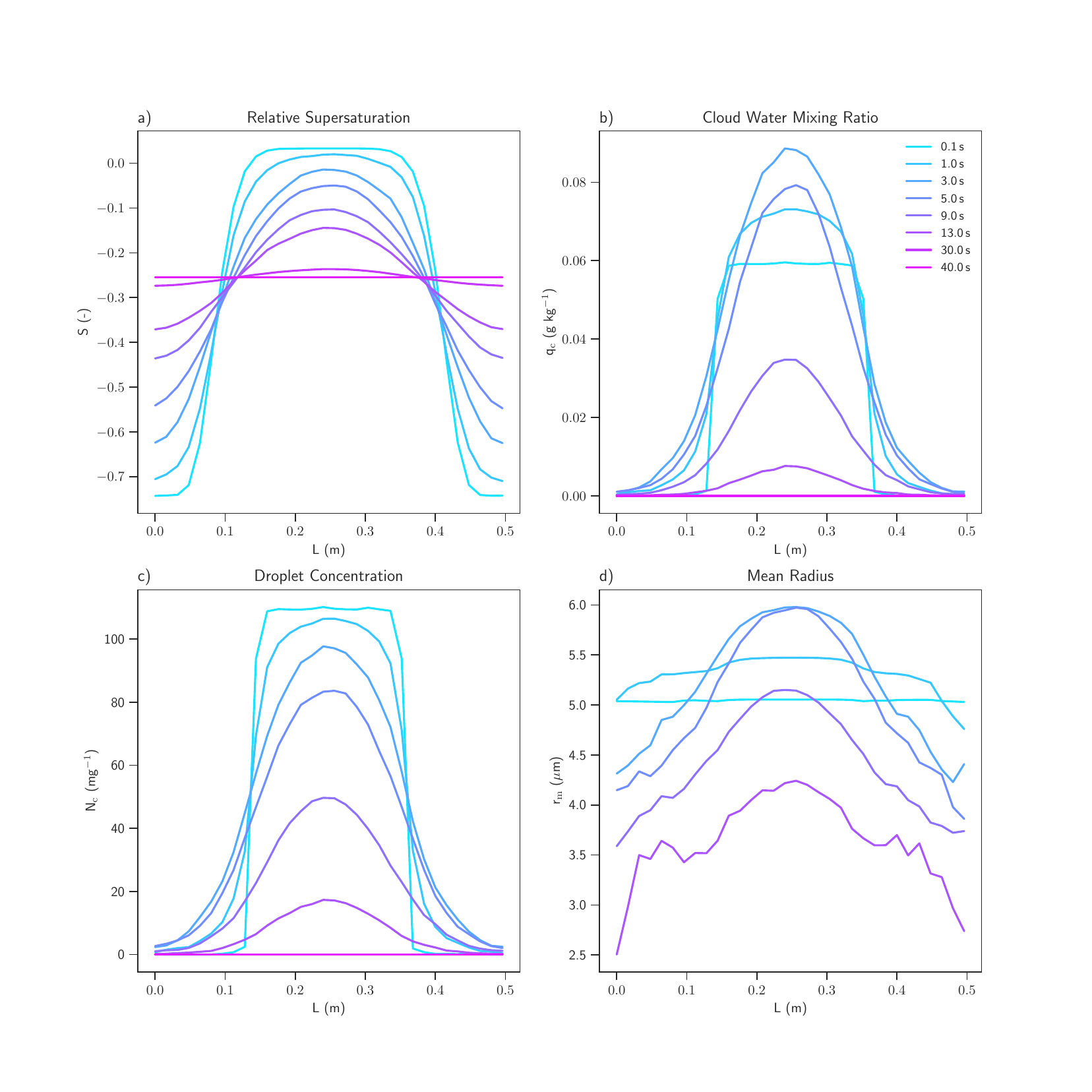}
    \caption{For individual timesteps (blue to green), the development of the relative supersaturation (a), the $q_{\mathrm{c}}$ (b), the droplet concentration (c) and the conditional average of the droplet radius (d) are shown along the entrainment profile. $r_{\mathrm{m}}$ is calculated as an arithmetic mean of an ensemble of 1000 simulation runs, where only gridboxes that contain at least one particle are taken into account.}
    \label{fig:dissprofiles}
\end{figure}

The qualitative behavior of the mixing process is demonstrated in Fig.\,\ref{fig:dissprofiles}, showing LEM results for $r_{\mathrm{i}} = 5\,\mu\text{m}$, $n_0$ thermodynamics, and $\varepsilon \,=\,1\,$cm$^2\,$s$^{-3}$. The initially supersaturated cloudy part (Fig.\,\ref{fig:dissprofiles}a) is depleted by cloud droplet condensation, resulting in a momentary increase in $q_\mathrm{c}$ and the average cloud droplet radius $r_{\mathrm{m}}$ (Figs.\,\ref{fig:dissprofiles}b and d). Note that our analysis defines cloud droplets as any particle with $r>0.5\,\mu\text{m}$. At the same time, however, the cloud is mixed with its subsaturated surroundings, which decreases $S$ and $N_\text{c}$ constantly (Figs.\,\ref{fig:dissprofiles}a and c). After about $3\,\text{s}$, net condensation switches to evaporation (Fig.\,\ref{fig:dissprofiles}b), and all droplets evaporate until the end of the simulation, at which a fully subsaturated domain establishes (Figs.\,\ref{fig:dissprofiles}a and b). Note that for more humid thermodynamic cases ($n_1$ to $n_3$), the domain saturates toward the end of the simulation, which prevents the full evaporation of all droplets (not shown).

\subsection{Sensitivity on Turbulence Intensity}
\label{sec:dissipation}

\begin{figure}[h!]
    \centering
    \includegraphics[width=\textwidth]{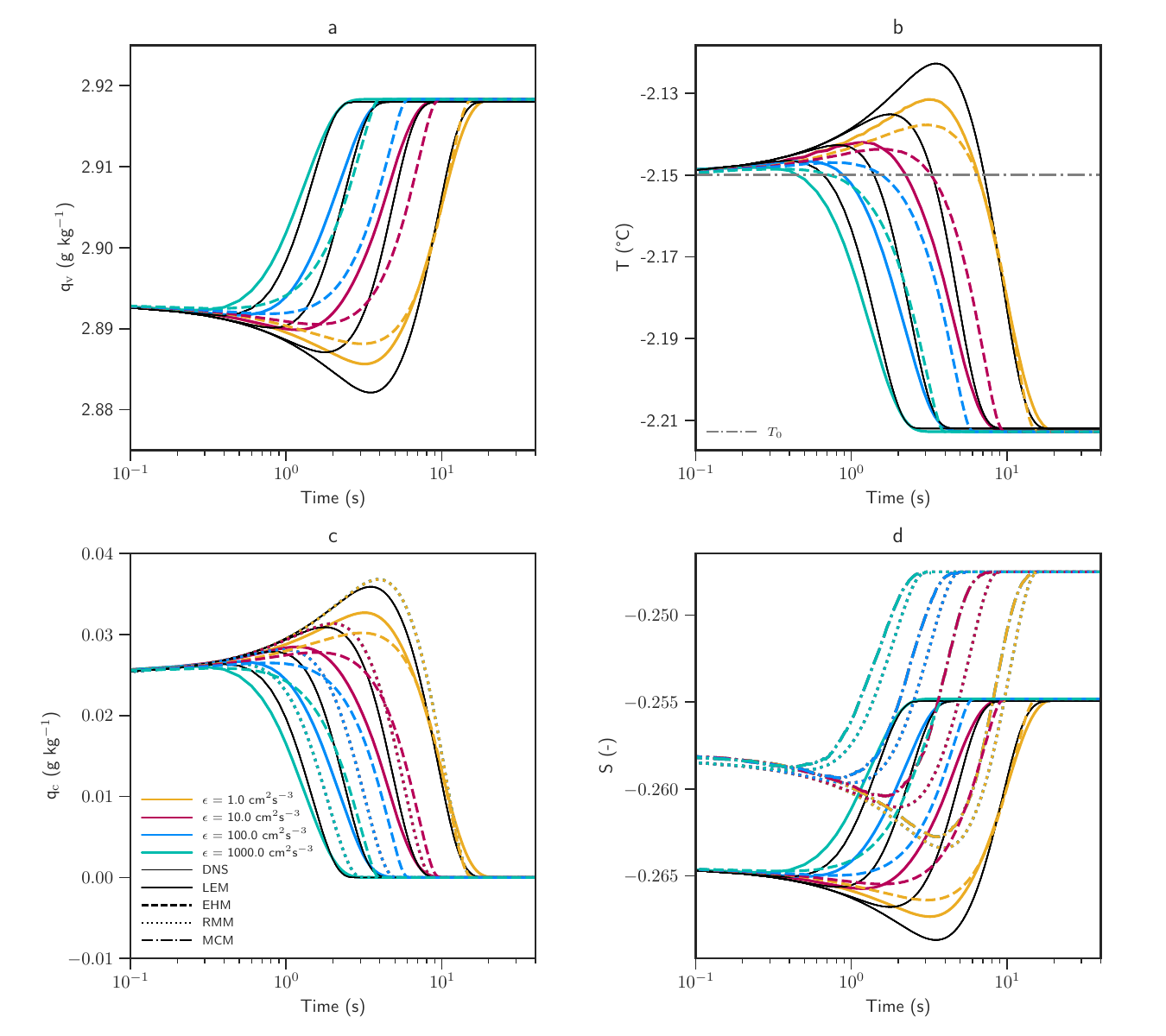}
    \caption{{Time evolution of the domain-averaged, that is, arithmetic mean of all grid boxes, (a) $q_\mathrm{v}$, (b) $T$, (c) $q_{\mathrm{c}}$ and (d) $S$, for four different energy dissipation rates (colors), and five models (pattern). Note that the DNS results are always in black to highlight them.}}
    \label{fig:dissstudythermodynamics}
\end{figure}

Now, we assess how changes in the degree of turbulence are captured by the models. Figures~\ref{fig:dissstudythermodynamics}a to d show the temporal evolution of the domain-averaged $q_\text{v}$, $T$, $q_\text{c}$, and $S$ for different $\varepsilon$, but the same $r_\text{i} = 5\,\mu\text{m}$ and $n_1$ thermodynamics. For readability, we omit the $\overline{(..)}$ to indicate domain averages in the following. Moreover, line colors indicate $\varepsilon$, and line patterns the analyzed models. Irrespective of $\varepsilon$, DNS data are shown by a thin black line to ease their identification. 

\begin{figure}[h!]
    \centering
    \includegraphics[width=\textwidth]{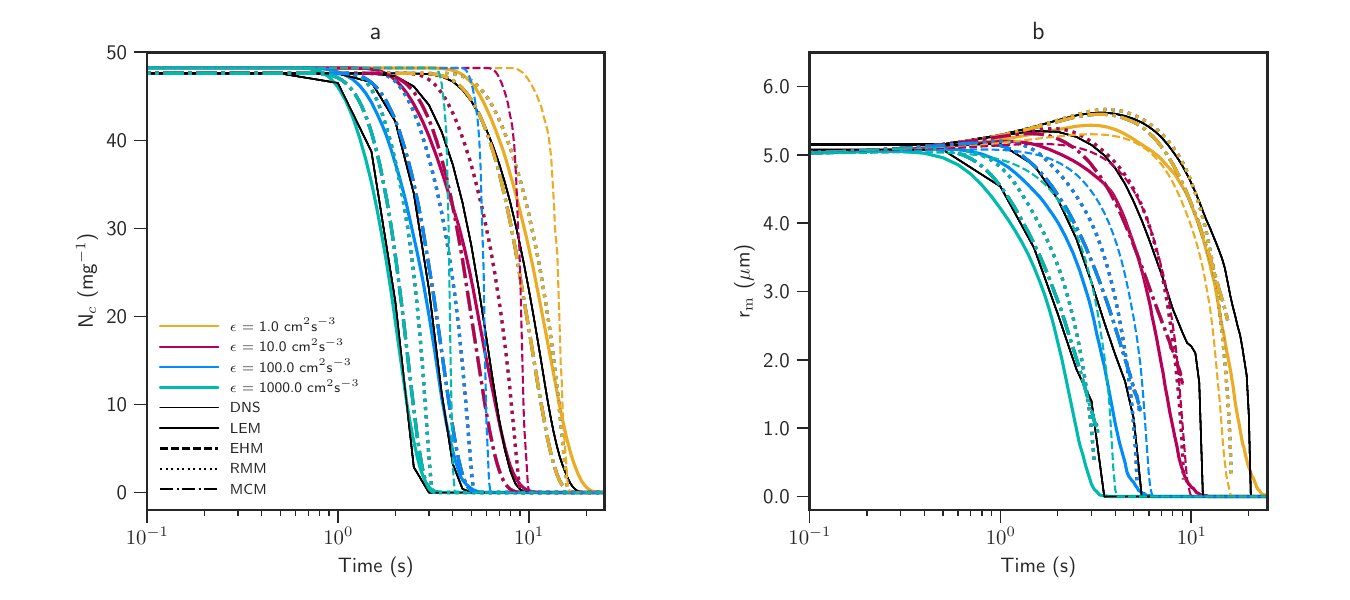}
    \caption{Time evolution of the domain-averaged, that is, arithmetic mean of all grid boxes, (a) $N_\mathrm{c}$, (b) $r_{\mathrm{m}}$ for four different energy dissipation rates (colors), and five models (pattern). Note that the DNS results are always in black to highlight them.}
    \label{fig:dissstudymicrophysics}
\end{figure}

As discussed for Fig.\,\ref{fig:dissprofiles}, the initially supersaturated cloudy part causes condensation during the onset of the simulation, resulting in a momentary reduction in $q_{\mathrm{v}}$ (Fig.\,\ref{fig:dissstudythermodynamics}a), a release of latent heat (Fig.\,\ref{fig:dissstudythermodynamics}b), an increase in $q_{\mathrm{c}}$ (Fig.\,\ref{fig:dissstudythermodynamics}c), and a decrease in $S$ (Fig.\,\ref{fig:dissstudythermodynamics}d). The initial condensation is followed by the complete evaporation of all droplets, and a domain void of $q_\text{c}$ at the end of the simulation (Fig.\,\ref{fig:dissstudythermodynamics}c). As for the initial condensation phase, the timescale for this process is determined by $\varepsilon$, which determines how quickly the cloud mixes with the ambient air. Specifically, the strongest $\varepsilon \,=\,1000\,$cm$^2\,$s$^{-3}$ leads to complete evaporation within $3\,\text{s}$, while the smallest $\varepsilon \,=\,1\,$cm$^2\,$s$^{-3}$ allows the cloud to survive for 20$\,$s.

Overall, the above-described behavior is captured by all models, with minor differences in the mixing rate. These are probably due to slight disagreements in the representation of a specific turbulence intensity, but much smaller than those caused by the analyzed variations in $\varepsilon$. Moreover, we see a distinct offset in $S$ predicted by RMM and MCM (Fig.\,\ref{fig:dissstudythermodynamics}d). Because DNS did not provide a domain-averaged $S$, $S$ is determined from the domain-averaged $q_\text{v}$ and $T$ for DNS, LEM, and EHM. On the other hand, RMM and MCM do not predict $q_\text{v}$ and $T$, but $S$ directly. Thus, the offset is a result of the nonlinear dependency of $S$ on $T$ that is not appropriately considered when a domain-averaged $T$ is used to determine the domain-averaged $S$. However, this discrepancy has no impact on the simulated physics, as the time series for $q_\text{c}$ suggests (Fig.\,\ref{fig:dissstudythermodynamics}c).

Figures\,\ref{fig:dissstudymicrophysics}a and b show the domain-averaged $N_{\mathrm{c}}$ and $r_{\mathrm{m}}$, respectively. While $r_{\mathrm{m}}$ increases during the initial condensation phase, $N_{\mathrm{c}}$ remains at its initial value. In the evaporation phase, all models show a decrease in $r_{\mathrm{m}}$, with full evaporation toward the end of the simulation. Interestingly, DNS, LEM, RMM, and MCM show a gradual decrease in $N_\text{c}$, while EHM predicts a constant $N_{\mathrm{c}}$ until  $r_{\mathrm{m}}$ reaches $0$. This behavior indicates that DNS, LEM, RMM, and MCM can represent inhomogeneous mixing, which is characterized by a decrease in $N_\text{c}$ due to the full evaporation of some droplets during the mixing process. EHM, however, seems to be biased toward homogeneous mixing, during which all droplets evaporate simulataneously, but none fully, resulting in a constant $N_\text{c}$ as long as not all droplets evaporate completely.

\subsection{Sensitivity on Ambient Humidity}

\begin{figure}[h!]
    \centering
    \includegraphics[width=\textwidth]{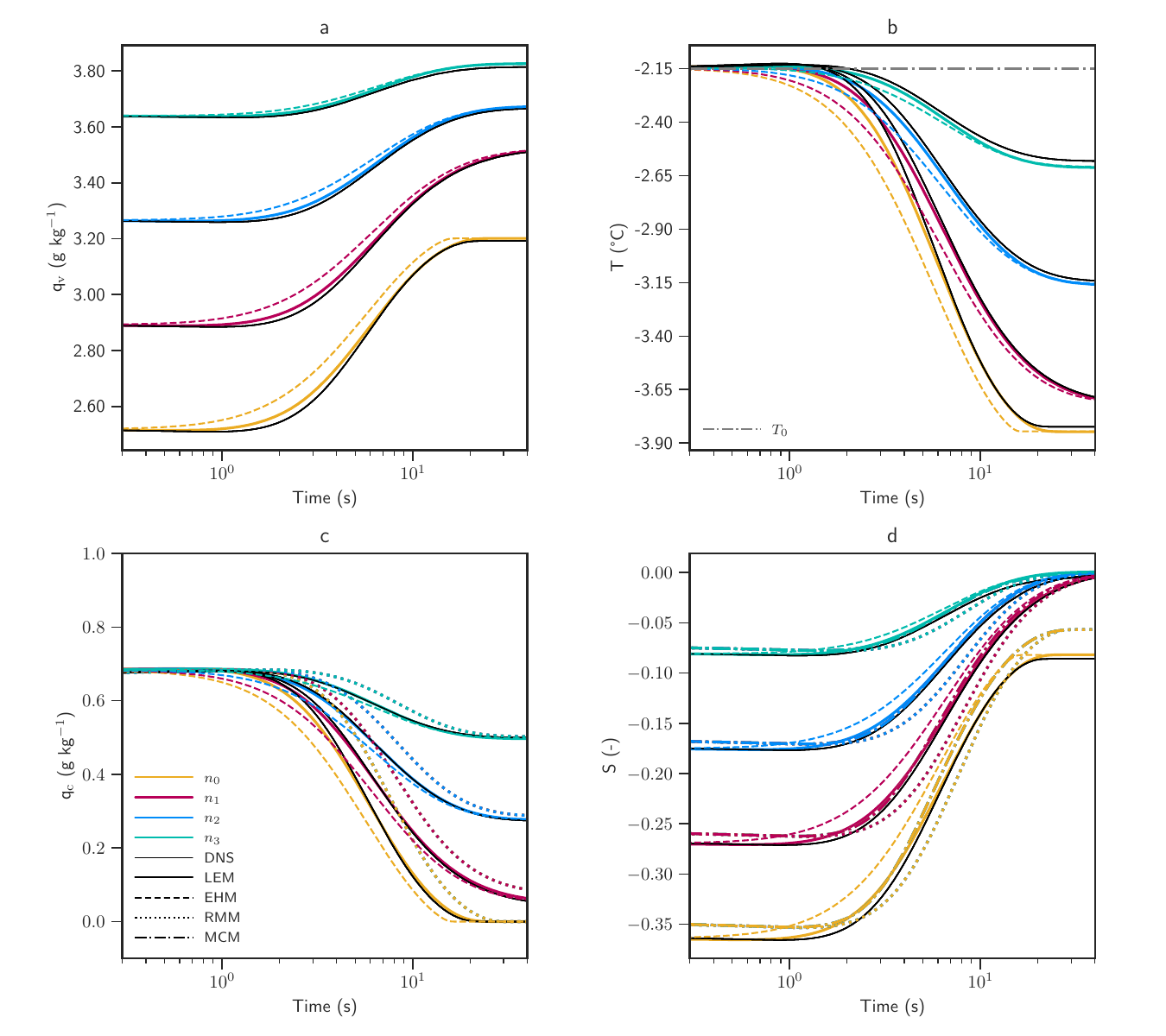}
    \caption{Time evolution of the domain-averaged, (a) $q_\mathrm{v}$, (b) $T$, (c) $q_{\mathrm{c}}$ and (d) $S$ for four different ambient humidities (colors), and five models (patterns). Note that the DNS results are always in black to highlight them. In (b) the reference temperature $T_0$ is shown (grey dashed-dotted line}
    \label{fig:humiditymicro}    
\end{figure}

\begin{figure}[h!]
    \centering
    \includegraphics[width=\textwidth]{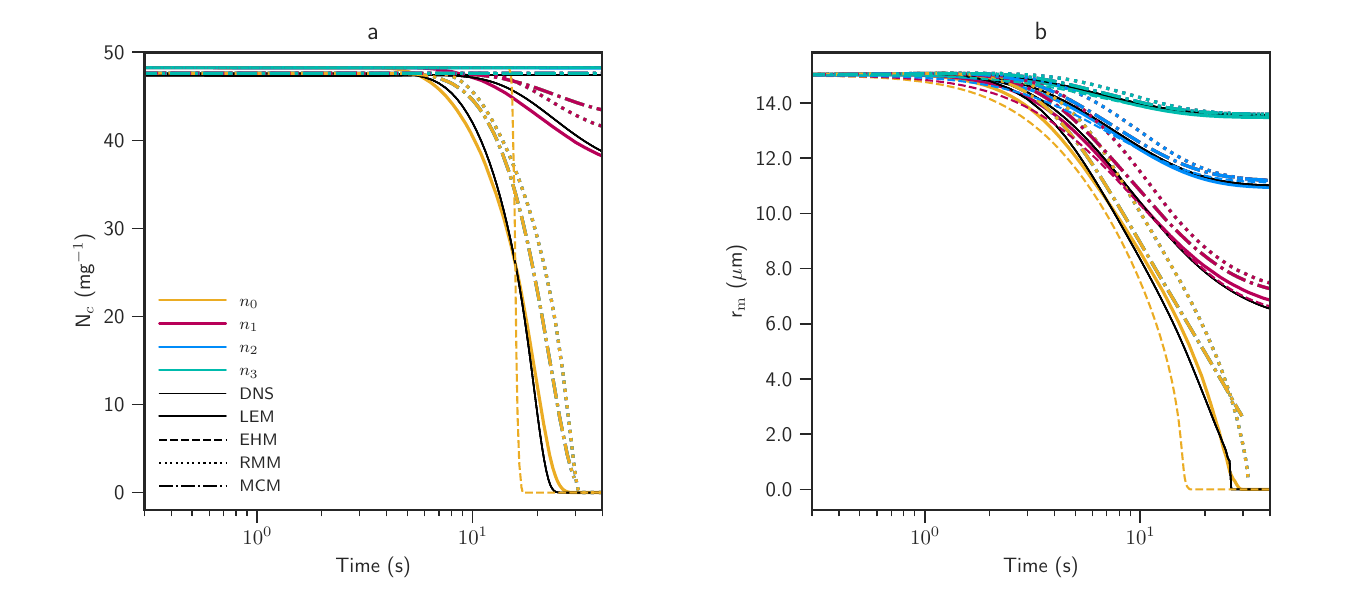}
    \caption{Time evolution of the domain-averaged, (a) $N_\mathrm{c}$, (b) $r_{\mathrm{m}}$, for four different ambient humidities (colors), and five models (patterns). Note that the DNS results are always in black to highlight them.}
    \label{fig:humiditythermo}
\end{figure}

Next, we test the susceptibility of all models to different ambient humidities, represented by the thermodynamic cases $n_0$ to $n_3$, all with $r_{\mathrm{i}} = 15\,\mu$m and $\varepsilon=33.75\,\text{cm}^{2}\,\text{s}^{-3}$. Particularly, we investigate cases in which the cloud does not fully evaporate by the end of the simulation (cases $n_1$ to $n_3$). As above, Fig.\,\ref{fig:humiditymicro} displays the temporal evolution of the domain-averaged $q_\text{v}$, $T$, $q_\text{c}$, and $S$. Again, all models capture the behavior well. Note, however, that differences among the models are less visible than for Sec.\,\ref{sec:dissipation} due to the larger range of values covered in the respective ordinates. Particularly, the above-discussed offset in $S$ from RMM and MCM is still present, as well as the same minor variability in the mixing rate. By the end of the mixing process, however, all models agree. This is not surprising since the end state is solely determined by thermodynamics, i.e., all models should reach the same state irrespective of cloud microphysics and dynamics. For RMM and MCM, however, this is only possible since the variability in $T$ is sufficiently small (Fig.\,\ref{fig:humiditymicro}b) to not violate the underlying assumption of a constant $T_0$ to determine $S$.

In Figs.\,\ref{fig:humiditythermo}a and b, the domain-averaged $N_{\mathrm{c}}$ and $r_{\mathrm{m}}$ are shown. While most models show again a very good agreement with the DNS, we still see that the EHM is not capable of capturing the gradual decrease in $N_\text{c}$ but rather represents an abrupt drop when all droplet evaporate. This effect is especially apparent for $n_0$ (orange lines) where the strongest evaporation takes place, but also for $n_1$ (red lines). As before, this indicates a bias toward homogeneous mixing in the EHM, while all other models indicate more inhomogeneous mixing. 

\subsection{Supersaturation Spectra and Droplet Size Distributions}

\begin{figure}
    \centering
    \includegraphics[scale=0.45]{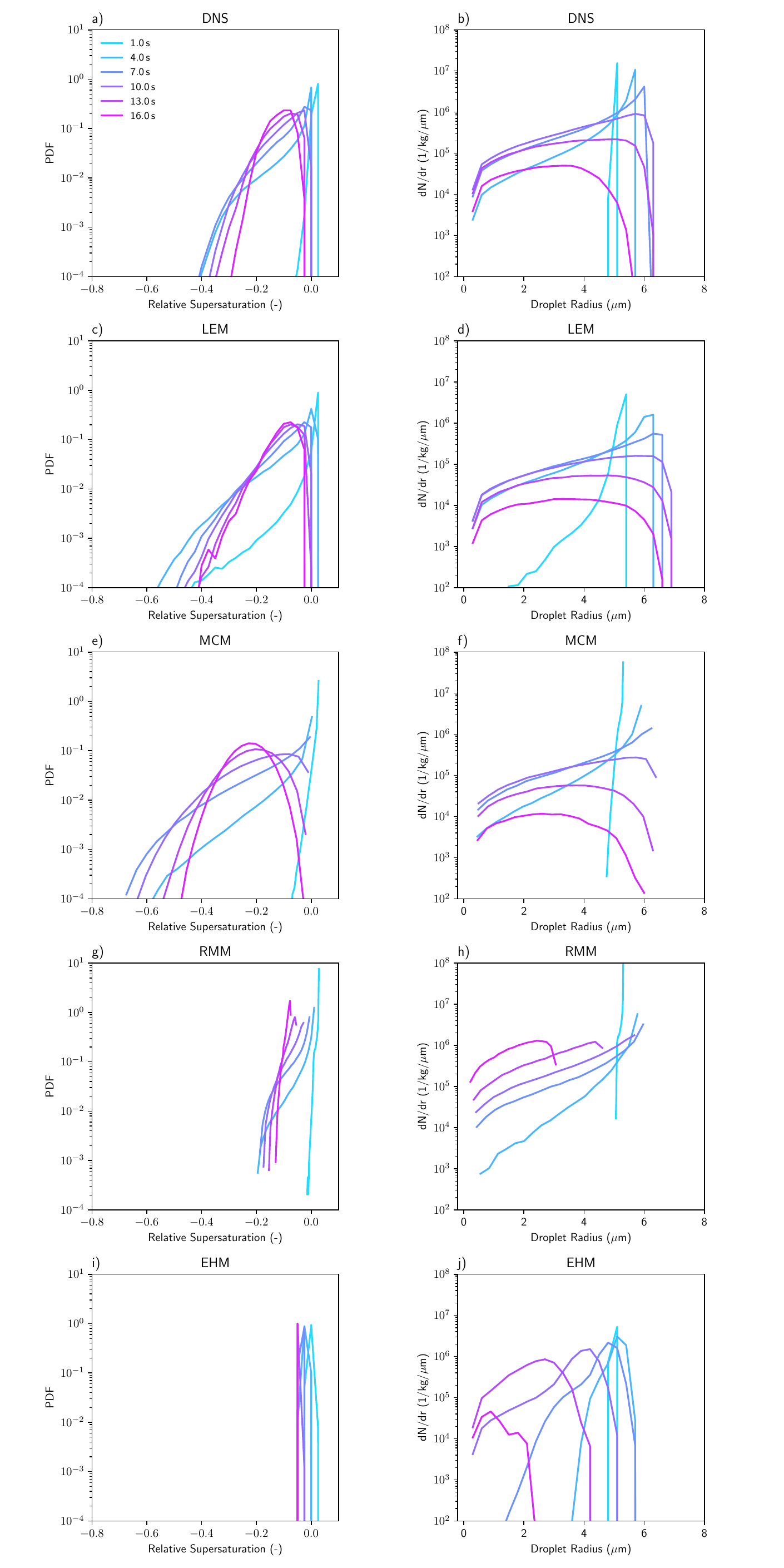}
    \caption{Time evolution of the supersaturation spectra (left column) and droplet size distributions (right column) with $n_1$ thermodynamics, $r_{\mathrm{i}}=5\,\mu$m, and $\varepsilon\,$=$\,1\,\mathrm{cm}^2\mathrm{s}^{-3}$. Different colors represent individual timesteps during the mixing process.}
    \label{fig:SATspectra_DSDspectra}
\end{figure}

\begin{figure}
    \centering
    \includegraphics[scale=0.45]{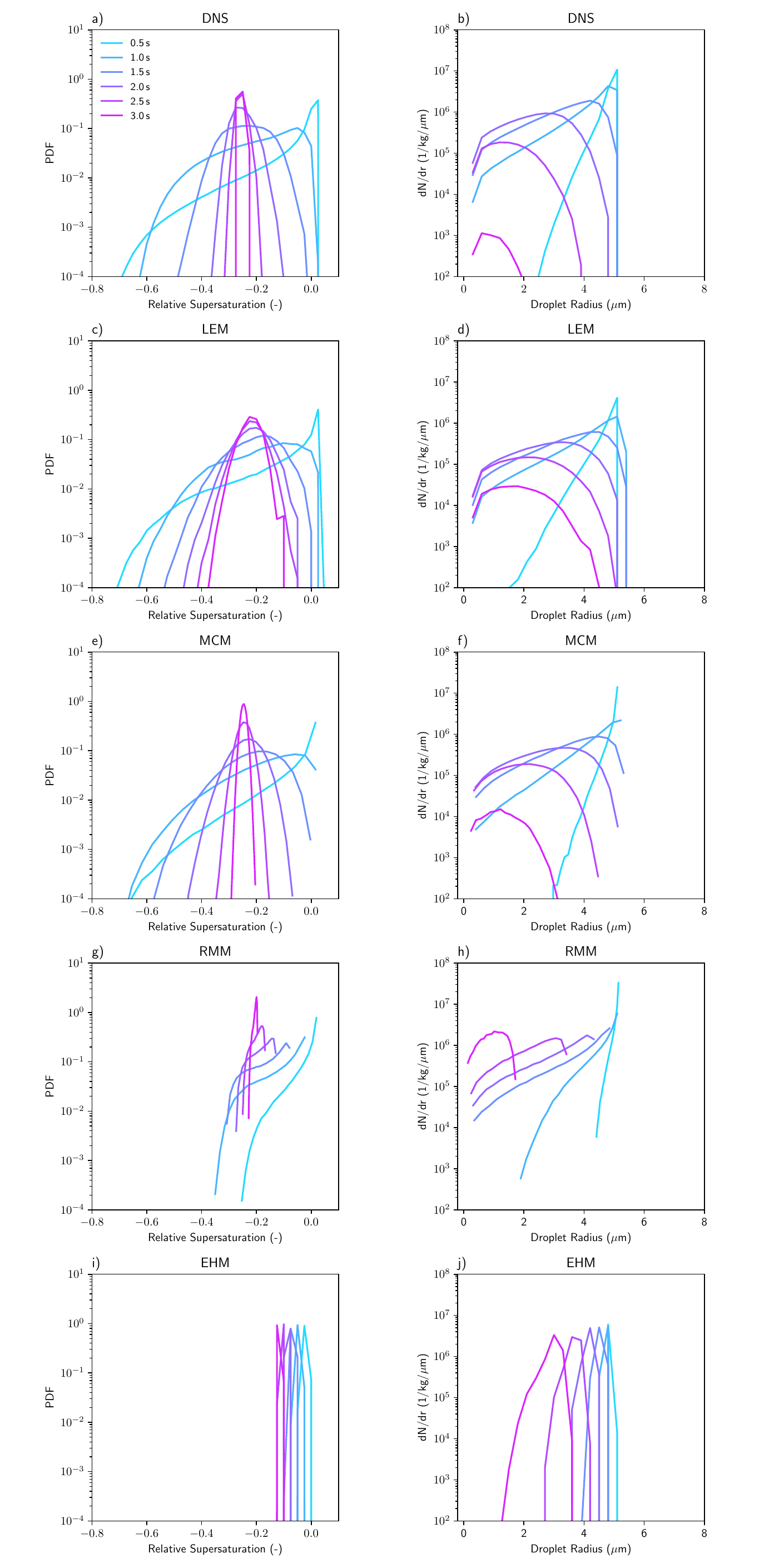}
    \caption{Time evolution of the supersaturation spectra (left column) and droplet size distributions (right column) with $n_1$ thermodynamics, $r_{\mathrm{i}}=5\,\mu$m, and $\varepsilon\,$=$\,1000\,\mathrm{cm}^2\mathrm{s}^{-3}$. Different colors represent individual timesteps during the mixing process.}
    \label{fig:SATspectra_DSDspectra1000}
\end{figure}

The ability of the presented models to represent the mixing process breaks down to their ability to accurately capture the development of $S$ and hence each droplets' growth history. For $\varepsilon = 1$ and $1000\, \mathrm{cm}^2 \,\mathrm{s}^{-3}$, the thermodynamics $n_1$, and $r_{\mathrm{i}}=5\,\mu $m, Lagrangian supersaturation spectra and the corresponding droplet size distributions are presented in Figs.\,\ref{fig:SATspectra_DSDspectra} and \ref{fig:SATspectra_DSDspectra1000}. Here, the term Lagrangian supersaturation refers to the supersaturations experienced by droplets. 

During the onset of the mixing process ($1\,\text{s}$), DNS and LEM show an asymmetric $S$ distribution, consisting of high $S$ from the cloud, and lower ambient $S$ experienced by those droplets already mixed outside of the cloud (Figs.\,\ref{fig:SATspectra_DSDspectra}a and c, and Figs.\,\ref{fig:SATspectra_DSDspectra1000}a and c). As expected, the mode of the $S$ distribution shifts to lower values as the mixing evolves. At the same time, the $S$ distribution also narrows, indicating the homogenization of cloudy and ambient air, resulting in a more Gaussian distribution of $S$. This behavior is especially visible for $\varepsilon = 1000\, \mathrm{cm}^2 \,\mathrm{s}^{-3}$ where mixing is stronger (Fig.\,\ref{fig:SATspectra_DSDspectra1000}), while the case with $\varepsilon = 1\, \mathrm{cm}^2 \,\mathrm{s}^{-3}$ maintains a broader and more asymmetric distribution throughout the mixing process (Fig.\,\ref{fig:SATspectra_DSDspectra}). As the MCM relaxes to a Gaussian $S$ distribution by design, it fails to represent the later $S$ distributions for $\varepsilon = 1\, \mathrm{cm}^2 \,\mathrm{s}^{-3}$ (Fig.\,\ref{fig:SATspectra_DSDspectra}e), while it agrees well with the DNS for $\varepsilon = 1000\, \mathrm{cm}^2 \,\mathrm{s}^{-3}$ (Fig.\,\ref{fig:SATspectra_DSDspectra1000}e). RMM and EHM largely fail to represent the broadening of the $S$ distribution (Figs.\,\ref{fig:SATspectra_DSDspectra}g and i, and Figs.\,\ref{fig:SATspectra_DSDspectra1000}g and i). While EHM assumes a very narrow $S$ distribution that slightly shifts to smaller values, RMM is at least able to represent some asymmetry in the $S$ distribution by including high $S$ from inside the cloud, but it misses representing the lowest ambient $S$ found in DNS. 
 
DNS, LEM, and MCM also agree well in their representation of the droplet size distribution, which contains large droplets from inside the cloud, and smaller droplets that have experienced some mixing and evaporation (Figs.\,\ref{fig:SATspectra_DSDspectra}b, d, and f, and Figs.\,\ref{fig:SATspectra_DSDspectra1000}b, d, and f). Thus, these models are able to represent inhomogeneous mixing during which some droplets evaporate, while others remain (almost) unblemished. The foundation for this is the ability of these models to represent a broad $S$ distribution (Figs.\,\ref{fig:SATspectra_DSDspectra}a, c, and e, and Figs.\,\ref{fig:SATspectra_DSDspectra1000}a, c, and e), as also recently shown by \citeA{lim2024life}. The RMM is able to capture this development of the droplet size distribution partially, although the largest droplets evaporate too fast (Figs.\,\ref{fig:SATspectra_DSDspectra}h and \ref{fig:SATspectra_DSDspectra1000}h). The EHM shows again a more homogeneous response, where the entire droplet size distribution evaporates to smaller sizes (Fig.\,\ref{fig:SATspectra_DSDspectra}j and Fig.\,\ref{fig:SATspectra_DSDspectra1000}j). Note that these differences among the models are especially strong for $\varepsilon = 1\, \mathrm{cm}^2 \,\mathrm{s}^{-3}$ (Fig.\,\ref{fig:SATspectra_DSDspectra}), which favor inhomogeneous mixing due to the low turbulence intensity, while the differences are smaller for $\varepsilon = 1000\, \mathrm{cm}^2 \,\mathrm{s}^{-3}$ (Fig.\,\ref{fig:SATspectra_DSDspectra1000}), which mixes more homogeneously.

\section{Discussion}

Extending our previous analysis, we now discuss some aspects of the individual models that need to be considered for their (potential) utilization as a subgrid-scale scheme. 

\subsection{``A Posteriori'' Nature of the EHM}
\label{sec:apost}

To represent the thermodynamic development correctly, the EHM needs to be tuned by adjusting the parameters $C_\text{EHM,1}$ and $C_\text{EHM,2}$ that steer the impact of mixing and cloud microphysics, as also noted by \citeA{Saito2021}. For this study, we determined $C_\text{EHM,1}=0.18$ and $C_\text{EHM,2}=0.03$ as useful tuning parameters for all assessed cases. However, different sets of tuning parameters might be necessary for more different environments that are encountered when the EHM is used as a subgrid-scale model in LES. Interactively adjusting $C_\text{EHM,1}$ and $C_\text{EHM,2}$ for a given environment is a challenge that needs to be addressed to further improve the applicability of the  EHM as a subgrid-scale model.

\subsection{LEM's Resolution Dependency}

To represent turbulence down to $\eta$, the LEM resolution is required to be smaller than $\eta/6$ \cite<e.g., >{Menon2011}.
Although the LEM requires substantially less computational resources than DNS, the high resolution required by the aforementioned constrain can make LEM relatively costly to operate. Thus, it is customary to apply the LEM at lower resolutions, for which $D_\kappa$ and $D_\text{v}$ are increased appropriately \cite<e.g.,>{krueger1997modeling}. However, a lower resolution limits the ability of the LEM to represent the details of small-scale mixing.

\begin{figure}[h!]
    \centering
    \includegraphics[width=\textwidth]{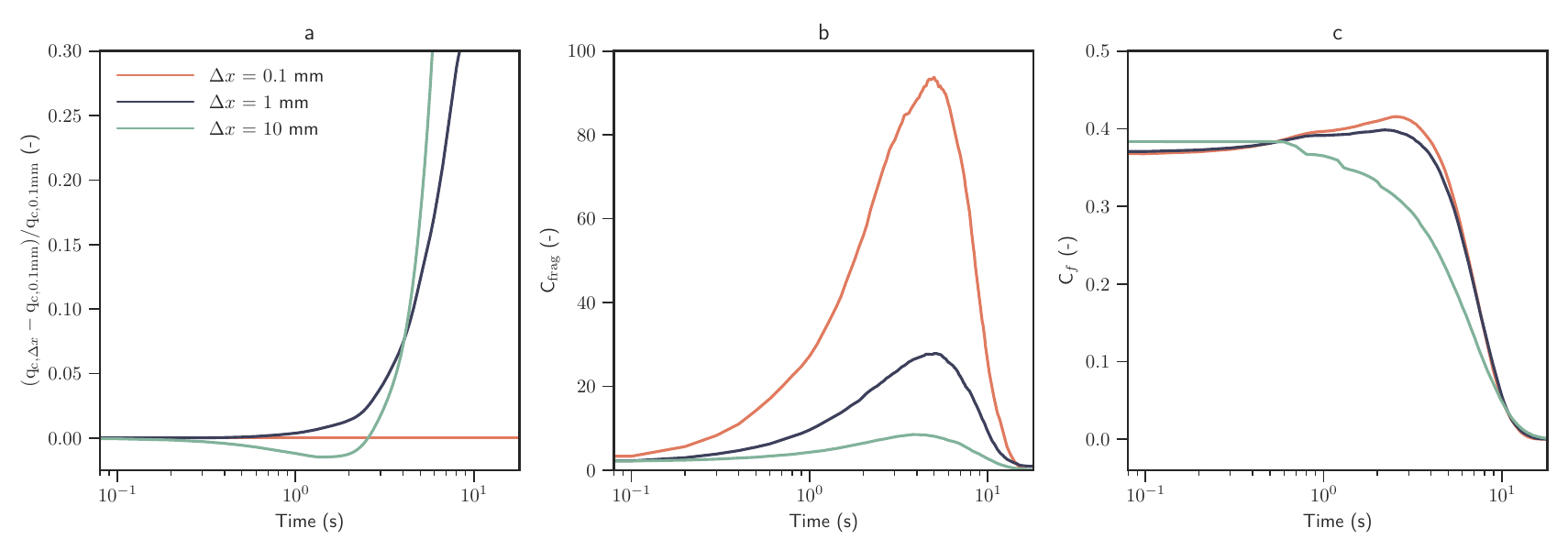}
    \caption{Time series of (a) the relative $q_\text{c}$ difference, (b) the fragmentation of the cloud $C_{\mathrm{frag}}$, and (c) the cloud fraction $C_\text{f}$  for different LEM resolutions (line colors).}
    \label{fig:LEMdiscussion}
\end{figure}

Figure~\ref{fig:LEMdiscussion} addresses some impacts when a lower LEM resolution is chosen for $\varepsilon = 1\, \mathrm{cm}^2 \,\mathrm{s}^{-3}$, $r_\text{i}= 5\,\mu \mathrm{m}$, and $n_1$ thermodynamics. The simulations are run with LEM grid spacings $\Delta x = 0.1$, $1.0$, and $10\,\text{mm}$. Figure~\ref{fig:LEMdiscussion}a shows the relative $q_{\mathrm{c}}$ difference from the highest resolution run, i.e., $(q_{\mathrm{c,}\Delta x}-q_{\mathrm{c,0.1\,mm}})/q_{\mathrm{c,0.1\,mm}}$, with the additional subscript indicating the resolution. One sees clearly that a larger grid spacing first decelerates initial condensation, resulting in a lower $q_\text{c}$, and then slows down the subsequent evaporation of the cloud, maintaining a higher $q_\text{c}$ (cf.\,Fig.\,\ref{fig:dissstudythermodynamics}c). The deceleration of initial condensation can be attributed to the faster decrease in cloud fraction $C_\text{f}$ at coarser resolutions, shown in Fig.\,\ref{fig:LEMdiscussion}c and determined from the fraction of grid boxes with $q_{\mathrm{c}}>10^{-2}\,\mathrm{g}\,\mathrm{kg}^{-1}$. Here, the larger grid spacing combined with the up-scaled $D_\kappa$ and $D_\text{v}$ enables ambient air to be transported deeper into the cloud, where it decelerates the initial condensation. At the same time, the higher $D_\kappa$ and $D_\text{v}$ artificially moisten the ambient air. This slows down the subsequent evaporation at coarser resolutions. This interpretation is supported by the higher cloud fragmentation $C_\text{frag}$ at higher resolutions, shown in Fig.\,\ref{fig:LEMdiscussion}b and determined from the number of patches with continuously cloudy grid boxes. (To avoid overestimating $C_\text{frag}$ at high resolutions, the LEM data is coarse-grained using the average distance between two cloud droplets inside the cloud part, about $2\,\text{mm}$.) While $C_\text{frag}$ remains small during the initial condensation, it quickly increases during evaporation at higher resolutions, indicating a highly intermittent cloud with cloudy patches exposed to (almost unprocessed) ambient air, leading to faster evaporation of some droplets at higher resolution. This analysis indicates that the ability of the LEM to represent inhomogeneous mixing depends on the resolution.

\subsection{The Closure-Dependency of MCM}

Although EHM, RMM, and MCM exhibit strong similarities in their formulation \cite<cf.>{Pope2011}, only MCM captures the development of cloud microphysics during the mixing successfully. To achieve this, MCM requires input from DNSs or other sources to consider the spatial variability in $S$, which is not considered in EHM and RMM. Thus, external data needs to be produced for various environments that are encountered during the (potential) application of MCM as a subgrid scale model. Here, machine learning could help to generate closures dependent on the environments experienced in a LES \cite<e.g.,>{Frezat2022,Jakhar2024}. A similar approach might be possible to estimate the tuning parameters for EHM in Sec.\,\ref{sec:apost}.

\section{Summary and Conclusion}

In this study, we assessed the ability of four statistical turbulence models (the LEM, EHM, RMM, and MCM) to represent the small-scale mixing of cloudy and cloud-free air by comparison with DNS data. 

The DNS is treated as ground truth, as it directly solves the underlying Navier-Stokes equations \cite{Kumar2014,Kumar2017}. The LEM employs a mapping approach to mimic how turbulence stretches and folds scalars, i.e., the thermodynamic quantities that drive the condensation and evaporation of cloud droplets during the mixing process \cite{Kerstein1988,krueger1993linear}. The EHM uses a spatially homogeneous Ornstein-Uhlenbeck process to determine the development of thermodynamics \cite{Pope1994,Grabowski2017}, while RMM is able to consider some spatial variability of thermodynamic quantities during the mixing process \cite{Pope2000,Fries2021}. The MCM uses a mapping closure to predict the non-Gaussian development of thermodynamics during mixing from Gaussian statistics \cite{Chen1989,Pope1991,Fries2023}. All models use a similar Lagrangian representation of cloud microphysics.

All statistical turbulence models accurately capture the development of thermodynamics during the mixing process. However, there are differences in the cloud microphysical properties. The differences are traced down to the ability of the statistical turbulence models to represent the supersaturation history experienced by the cloud droplets. We showed that this history is closely linked to the spatial variability of the supersaturation during the mixing process, which contains subsaturations from ambient air as well as supersaturations from the cloud. Only if this broad distribution of supersaturations and its development is represented in a model, it is able to represent inhomogeneous mixing during which some droplets evaporate completely while others are largely unaffected \cite{Baker1979,lim2024life}. This is the case for LEM, MCM, and partially RMM. The EHM, however, fails to represent this aspect of the mixing process, as all droplets experience approximately the same subsaturation. The resultant constant droplet number concentration can be interpreted as homogeneous mixing.

We have demonstrated that simpler and more computationally efficient models can also capture certain aspects of small-scale cloud-edge mixing. Nevertheless, DNS remains essential for accurately resolving the intricate physical processes that are beyond the scope of these models. However, the applicability of LEM, EHM, RMM, and MCM as subgrid-scale models in LES needs to be evaluated further. LEM and EHM are already used as subgrid-scale models to represent the effects of turbulent supersaturation fluctuations on droplet growth in LESs \cite{HoffmannFeingold2019,Chandrakar2021}. However, for the use of MCM as a subgrid-scale model, some problems need to be solved (e.g., the generation of highly scenario-specific closure data). Nonetheless, the statistical turbulence models analyzed here constitute practicable means to bridge the gap between small-scale turbulence and clouds, and thus constitute a path forward to assess the role of clouds in the climate system in a more holistic framework.

\section{Open Research}

The model data for this study can be obtained from \citeA{kainz_2024_13813360}. The models are available from the respective authors upon request.

\acknowledgments

J.\,Kainz and F.\,Hoffmann were supported by the Emmy Noether program of the German Research Foundation (DFG) under Grant HO~6588/1-1. G.\,Sardina  acknowledges support from Vetenskapsr\aa{}det (VR) under grant no.\,2022-03939, and his computations were enabled by resources provided by the National Academic Infrastructure for Supercomputing in Sweden (NAISS), partially funded by the Swedish Research Council through grant agreement no. 2022-06725. B.\,Mehlig was supported by VR grant no. 2021-4452. The DNS work by B.\,Kumar,  N.\,N.\,Makwana, and S.\,Ravichandran has been carried out using High Performance Computing system (HPCS) facility available at IITM. The IITM is fully funded by the Ministry of Earth Sciences, Government of India. SR is also supported through the Monsoon Mission project IITM/MM-III/2023/IND-4.

\bibliography{bibliography}

\end{document}